%% file: main.tex
\documentclass[conference]{IEEEtran}
\IEEEoverridecommandlockouts
\usepackage{multirow}
\usepackage{amsmath,amsfonts}
\usepackage{array}
\usepackage{textcomp}
\usepackage{booktabs}
\usepackage{tabularx}
\usepackage{url}
\usepackage{verbatim}
\usepackage{graphicx}
\usepackage{cite}
\usepackage[table,xcdraw]{xcolor}
\usepackage[printonlyused, nolist]{acronym}
\usepackage{tikz}
\usepackage{placeins}
\usepackage{stfloats}

\newcommand\copyrighttext{%
  \footnotesize \textcopyright \the\year{} IEEE. Personal use of this material is permitted. Permission from IEEE must be obtained for all other uses, including reprinting/republishing this material for advertising or promotional purposes, creating new collective works for resale or redistribution to servers or lists, or reuse of any copyrighted component of this work in other works.}

\newcommand\copyrightnotice{%
\begin{tikzpicture}[remember picture,overlay]
\node[anchor=north,yshift=0pt] at (current page.north) {\fbox{\parbox{\dimexpr0.9\textwidth-\fboxsep-\fboxrule\relax}{\copyrighttext}}};
\end{tikzpicture}%
}

\begin{document}
\input{acronyms}



\title{HELENA for 5G NR LEO NTN Channel Estimation: A Comparative Evaluation
\thanks{This work was supported by the imec.icon RAPIDNESS project,
co-financed by imec and Flanders Innovation \& Entrepreneurship
(HBC.2024.0772).}
}
\author{
\IEEEauthorblockN{
Miguel Camelo Botero,
Nina Slamnik-Kriještorac,
and Johann Marquez-Barja
}
\IEEEauthorblockA{
\textit{IDLab, University of Antwerp -- imec}\\
Antwerp, Belgium
}
}

\maketitle

\copyrightnotice

\begin{abstract}
\ac{DL}-based channel estimation has shown high accuracy and low latency in terrestrial \acs{5G-NR}, but \ac{LEO} \acp{NTN} introduce Doppler and synchronization impairments that may require NTN-specific architectures. We investigate whether \acs{HELENA}, a dual-attention neural channel estimator originally designed for terrestrial channels, remains effective after\ac{NTN} retraining and suitable across high-performance and power-constrained inference platforms. Its unchanged architecture is trained on paired receiver-compensated and residual-impaired datasets and evaluated alongside terrestrial-origin, NTN-specific, and classical estimators. \acs{HELENA}'s SNR-averaged \ac{NMSE}, computed by arithmetic averaging in linear scale, remains within 0.29 and 0.72~dB of the most accurate learned estimator, \acs{AttRNet}, in the two conditions, respectively. It outperforms global \ac{LMMSE} under compensation and all evaluated \acs{LMMSE} variants under residual impairments, although profile-aware filtering achieves lower error under compensation. On an RTX PRO 4500, \acs{FP16} TensorRT inference reaches 0.0573~ms \ac{P99} latency, below the adopted 0.5~ms estimator budget. On a 10~W Jetson Orin NX, \acs{HELENA} consumes 66.0\% less energy per sample than \acs{AttRNet}, but its 0.9620~ms \acs{P99} exceeds the budget; only the less accurate \acs{LSiDNN}-48 meets it. Thus, \acs{HELENA} remains effective without NTN-specific redesign for the evaluated task, while accurate embedded
inference remains a latency challenge.
\end{abstract}


\acresetall
\input{sections/section1}

\input{sections/section2}
\input{sections/section3}
\input{sections/section5}
\input{sections/section6}
\input{sections/section7}

\input{sections/section8}

\bibliographystyle{IEEEtran}
\bibliography{main.bib}

\end{document}

%% file: acronyms.tex
\begin{acronym}
    \acro{AI}{Artificial Intelligence}
    \acro{ML}{Machine Learning}
    \acro{CNN}{Convolutional Neural Network}
    \acro{FC}{Fully Connected Layers}
    \acro{5G-NR}[5G NR]{Fifth-generation New Radio}
    \acro{5G}{Fifth-Generation}
    \acro{DL}{Deep Learning}
    \acro{ReLU}{Rectified Linear Unit}
    \acro{DNN}{Deep Neural Network}
    \acro{ONNX}{Open Neural Network Exchange}
    \acro{GPU}{Graphics Processing Unit}
    \acro{CPU}{Central Processing Unit}
    \acro{SRCNN}{Super Resolution Convolutional Neural Network}
    \acro{EDSR}{Enhanced Deep Super-Resolution}
    \acro{ViT}{Vision Transformer}
    \acro{SR}{Super-Resolution}
    \acro{IR}{Image Restoration}
    \acro{LR}{Low-Resolution}
    \acro{LS}{Least Squares}
    \acro{LI}{Linear Interpolation}
    \acro{HR}{High-Resolution}
    \acro{MDSR}{Multi-Scale Approach}
    \acro{MAE}{Mean Absolute Error}
    \acro{CE}{Channel Estimation}
    \acro{NMSE}{Normalized Mean Squared Error}
    \acro{MSE}{Mean Squared Error}
    \acro{MHSA}{Multi-Head Self-Attention}
    \acro{BER}{Bit Error Rate}
    \acro{CSI}{Channel State Information}
    \acro{3GPP}{3rd Generation Partnership Project}
    \acro{OFDM}{Orthogonal Frequency-Division Multiplexing}
    \acro{SNR}{Signal-to-Noise Ratio}
    \acro{AWGN}{Additive White Gaussian Noise}
    \acro{SISO}{Single-Input Single-Output}
    \acro{I/Q}{In-Phase and Quadrature Components}
    \acro{1D}{One-Dimensional}
    \acro{2D}{Two-Dimensional}
    \acro{AR}{Augmented Reality}
    \acro{VR}{Virtual Reality}
    \acro{UHD}{Ultra-High-Definition}
    \acro{TDL}{Tapped Delay Line}
    \acro{TTI}{Transmission Time Interval}
    \acro{DnCNN}{Denoising Convolutional Neural Network}
    \acro{Conv}{Convolutional}
    \acro{NR}{New Radio}
    \acro{MMSE}{Minimum Mean Squared Error}
    \acro{LMMSE}{Linear Minimum Mean Squared Error}
    \acro{DRN}{Deep Residual Networks}
    \acro{ResNet}{Residual Network}
    \acro{RB}{Resource Block}
    \acro{SCS}{Subcarrier Spacing}
    \acro{CP}{Cyclic Prefix}
    \acro{PDSCH}{Physical Downlink Shared Channel}
    \acro{SR-Net}{Super-Resolution Network}
    \acro{CIR}{Channel Impulse Response} 
    \acro{SE}{Squeeze-and-Excitation}
    \acro{ECA}{Efficient Channel Attention}
    \acro{GAP}{Global Average Pooling}
    \acro{SRDnNet}{Super Resolution De-noising Convolutional Neural Network}
    \acro{ReEsNet}{Residual channel Estimation Network}
    \acro{UE}{User Equipment}
    \acroplural{UE}[UEs]{User Equipment Devices}
    \acro{IQ}{In-Phase and Quadrature}
    \acro{HELENA}{High-Efficiency Learning-based channel Estimation using dual Neural Attention}
    \acro{FLOPS}{Floating-Point Operations per Second}
    \acro{Conv2D}{2D Convolutional}
    \acro{MBConv}{Mobile Inverted Bottleneck Convolution}
    \acro{HARQ}{Hybrid Automatic Repeat Request}
    \acro{FPGA}{Field-Programmable Gate Array}
    \acro{LSiDNN}{LS-augmented interpolated Deep Neural
Network}
    \acro{AttRNet}{Attention mechanism and Residual
Network}
    \acro{CEViT}[CE-ViT]{Channel Estimator Vision Transformer}
    \acro{PD-CEViT}{Pilot-Design Channel Estimator Vision Transformer}
    \acro{OTFS}{Orthogonal Time Frequency Space}
    \acro{UAV}{Unmanned Aerial Vehicle}
    \acro{LoS}{Line-of-Sight}
    \acro{NLoS}{Non-Line-of-Sight}
    \acro{NDT}{Network Digital Twin}
    \acro{MIMO}{Multiple-Input Multiple-Output}
    \acro{dB}{Decibels}
    \acro{ProEsNet}{Progressive Estimation Network}
    \acro{EPformer}{Efficient Parallel Transformer}
    \acro{NTN}{Non-Terrestrial Network}
    \acro{TN}{Terrestrial Network}
    \acro{LEO}{Low Earth Orbit}
    \acro{NGSO}{Non-Geostationary Satellite Orbit}
    \acro{CFO}{Carrier-Frequency Offset}
    \acro{DM-RS}{Demodulation Reference Signal}
    \acro{PA-LS}{Pilot-Aided Least Squares}
    \acro{DA-LS}{Data-Aided Least Squares}
    \acro{MDELAN}{Multi-Dilated Efficient Layer Aggregation Network}
    \acro{TPE}{Tree-Structured Parzen Estimator}
    \acro{P99}{99th-Percentile}
    \acro{FP16}{16-bit Floating Point}
    \acro{FP32}{32-bit Floating Point}
    \acro{INT8}{8-bit Integer}
    \acro{GNSS}{Global Navigation Satellite System}
    \acro{SIB19}{System Information Block 19}
    \acro{ICI}{Inter-Carrier Interference}
    \acro{RE}{Resource Element}
    \acro{CDM}{Code-Division Multiplexing}
    \acro{QPSK}{Quadrature Phase-Shift Keying}
    \acro{SSB}{Synchronization Signal Block}
    \acro{PBCH}{Physical Broadcast Channel}
    \acro{BLER}{Block Error Rate}
    \acro{COTS}{Commercial Off-the-Shelf}
    \acro{FLOPs}{Floating-Point Operations}
\end{acronym}

%% file: sections/section1.tex
\bstctlcite{IEEEexample:BSTcontrol}
\section{Introduction}\label{sec:introduction}

\IEEEPARstart{F}{ifth-generation} \acl{NR} (\acs{5G-NR})\acused{5G-NR}\acused{NR} \acp{NTN} use satellites or airborne platforms to extend cellular access to remote, maritime, and aeronautical areas. In \ac{LEO} access, orbital motion creates large Doppler shifts and time-varying propagation delays, while \acp{UE} at different positions within a satellite beam experience different Doppler and delay conditions~\cite{TR38811}. These impairments complicate synchronization and \ac{CE} compared to terrestrial \ac{5G-NR}.

To mitigate Doppler, the transmitter can pre-compensate the component common to all users in a beam using satellite ephemeris and a beam-reference position~\cite{Meshram2025InitialSync}. Nevertheless, users away from the reference position experience residual frequency offsets, while mobility introduces additional channel variation. Assisted \ac{NR} \ac{NTN} receivers use \ac{SIB19} assistance information, \ac{GNSS} positioning, \ac{CP} redundancy, and reference signals to estimate and correct these impairments~\cite{TS38331,TR38821,vanDeBeek1997,Chen2025NTNDoppler}. Residual synchronization errors still introduce phase distortion and \ac{ICI}, such that the channel estimator must recover the effective post-processing channel rather than only the underlying multipath channel typically considered in terrestrial evaluations~\cite{soltani2019deep,HELENA2025}. 

This requirement raises the question of whether existing channel-estimation approaches can remain effective under \ac{NTN} conditions. Pilot-aided \ac{LS} estimation is computationally inexpensive but highly sensitive to noise, whereas \ac{LMMSE} improves robustness using second-order channel statistics at the cost of increased acquisition complexity and computational overhead~\cite{lee2008training,Edfors1998}. Learning-based alternatives include convolutional \ac{SR} and residual networks~\cite{soltani2019deep,maruyama2021,Gao2025,Zhang2024}, fully connected architectures~\cite{Sharma2024}, transformers~\cite{CEViT2023}, and the \ac{CE} components of model-driven receivers such as MDX~\cite{MDX2025}. More recently, \ac{MDELAN}~\cite{MDELAN2026} extended MDX to \acp{NTN} with a dedicated focus on channel estimation. 

While these studies demonstrate the potential of \ac{DL}-based \ac{CE}, it remains unclear whether efficient terrestrial architectures require \ac{NTN}-specific redesigns or can retain their performance after retraining on \ac{NTN} data. In this context, our previous work \ac{HELENA}, originally developed for terrestrial \ac{OFDM} \ac{CE}~\cite{HELENA2025}, is particularly attractive because it combines sparse pilot inputs with low computational complexity. However, its accuracy--efficiency trade-off under \ac{NTN} impairments, especially across different residual-Doppler conditions and deployment platforms, has not yet been assessed. 

To address this gap, we retrain \acs{HELENA} without architectural modification and compare it against two \acs{HELENA} variants, seven terrestrial \ac{DL} estimators, and one \ac{NTN}-specific estimator, all trained on the same \ac{NTN} datasets. The \acs{HELENA} variants isolate the impact of its second attention mechanism and reduced attention capacity. Paired receiver-compensated (\acs{NTN}-1) and residual-impaired (\acs{NTN}-2) datasets enable the assessment of receiver Doppler-compensation effects, while \ac{FP16} deployment on both high-performance and embedded accelerators evaluates latency and energy efficiency.

Building on this evaluation framework, the main contributions of this paper are as follows:
\begin{enumerate}
    \item[a)] Paired receiver-compensated (\acs{NTN}-1) and residual-impaired (\acs{NTN}-2) datasets isolating receiver Doppler-compensation effects.
    \item[b)] A common comparison of eleven learned configurations, \acs{HELENA} and two variants, seven other terrestrial estimators, and one \ac{NTN}-specific estimator, with practical pilot-based and statistical baselines under both receiver conditions.
    \item[c)] An accuracy--complexity--latency--energy evaluation using \ac{FP16} TensorRT on high-performance and power-constrained embedded accelerators.
\end{enumerate}

The paper continues as follows: Sections~\ref{sec:system}--\ref{sec:dataset} present the system model, compared estimators, and datasets. Section~\ref{results} reports the results, Section~\ref{sec:discussion} discusses their implications, and Section~\ref{sec:conclusion} concludes the paper.

%% file: sections/section2.tex
\section{System Model and Problem Statement}\label{sec:system}

We consider a downlink \ac{SISO} \ac{5G-NR} \ac{OFDM} link between a \ac{LEO} satellite and a terrestrial \ac{UE}. Let $f_d(\theta)$ denote the signed satellite-induced Doppler shift, in hertz, observed at a ground location with satellite elevation angle $\theta$. In Fig.~\ref{fig:ntn_system_model}, the beam-center reference (BC) is located at the footprint center, while the \ac{UE} may occupy another point in the beam. Their elevation angles $\theta_{\mathrm{BC}}$ and $\theta_{\mathrm{UE}}$ define the beam-common Doppler $f_{d,\mathrm{common}}=f_d(\theta_{\mathrm{BC}})$ and the satellite Doppler at the \ac{UE} $f_{d,\mathrm{sat}}=f_d(\theta_{\mathrm{UE}})$, respectively. The transmitter is assumed to know $f_{d,\mathrm{common}}$ from the satellite and beam-reference geometry and removes it before radiation. The residual \ac{CFO} at the receiver input is then
\begin{equation}
f_{d,\mathrm{pre}}=f_{d,\mathrm{sat}}-f_{d,\mathrm{common}},
\label{eq:residual_doppler_pre}
\end{equation}
where \ac{UE} mobility is represented by the maximum Doppler shift of the time-varying multipath channel. The subscript $d$ denotes Doppler, while $\mathrm{pre}$ denotes the receiver-input offset after transmitter pre-compensation but before receiver processing. Equivalently, transmitter pre-compensation maps the complex baseband waveform $x(t)$ into $x_{\mathrm{pc}}(t)=x(t)e^{-j2\pi f_{d,\mathrm{common}}t}$, where $t$ denotes time in seconds and $j=\sqrt{-1}$. The label $\mathrm{pc}$ denotes transmitter pre-compensation.

To illustrate how the residual offsets appear on the \ac{OFDM} grid, let $i$ index subcarriers and $k$ index \ac{OFDM} symbols. The corresponding baseband subcarrier frequency and symbol reference time are $f_i$ in hertz and $t_k$ in seconds. A diagonal approximation of the received \ac{RE} before receiver compensation is
\begin{equation}
Y^{\mathrm{pre}}_{i,k}=H^{\mathrm{can}}_{i,k}X_{i,k}
e^{j2\pi f_{d,\mathrm{pre}}t_k}e^{-j2\pi f_i\tau}+Z_{i,k},
\label{eq:received_ntn}
\end{equation}
where $X_{i,k}$ and $Y^{\mathrm{pre}}_{i,k}$ are the transmitted and received grid symbols, respectively. $H^{\mathrm{can}}_{i,k}$ is the canonical ($\mathrm{can}$), noise-free multipath response in the transmitter reference frame, excluding the residual Doppler and timing transformations represented by the exponential terms. It retains the channel time variation caused by \ac{UE} mobility. Moreover, $\tau$ is the timing error in seconds relative to the nominal \ac{OFDM} demodulation window before receiver timing alignment, and $Z_{i,k}\sim\mathcal{CN}(0,\sigma^2)$ is circularly symmetric complex Gaussian noise with total complex variance $\sigma^2=\mathbb{E}[|Z_{i,k}|^2]$ per \ac{RE}.

The exponentials describe Doppler-induced phase evolution across symbols and timing-induced phase slope across subcarriers. Equation~\eqref{eq:received_ntn} omits \ac{ICI}; the waveform-level simulations in Section~\ref{sec:dataset} retain it through channel filtering, frequency compensation, and timing alignment before \ac{OFDM} demodulation. 

We now turn to the receiver processing considered in this work. We use $s$ to identify the two receiver-processing configurations used in this paper: $s=1$ denotes the compensated \acs{NTN}-1 condition and $s=2$ the residual-impaired \acs{NTN}-2 condition. For configuration $s$, let $\hat\tau_s$ and $\hat f_{d,\mathrm{RX},s}$ denote the applied timing and residual-Doppler estimates, in seconds and hertz, respectively. A hat denotes an estimated quantity; $\mathrm{RX}$ labels receiver-side estimation and $\mathrm{post}$ labels the offsets remaining after receiver processing. The remaining offsets after receiver processing are
\begin{equation}
\tau_{\mathrm{post},s}=\tau-\hat\tau_s,\qquad
f_{d,\mathrm{post},s}=f_{d,\mathrm{pre}}-\hat f_{d,\mathrm{RX},s}.
\label{eq:post_rx_residuals}
\end{equation}
Thus, transmitter pre-compensation removes the beam-common satellite component, whereas receiver-side compensation estimates and removes the remaining location-dependent frequency offset. Bold uppercase symbols denote matrices, including resource grids, while bold lowercase symbols denote vectors. Let $\mathbf y$ denote the noisy received time-domain waveform and let $\mathbf y^{\mathrm{clean}}$ denote its matched counterpart, generated with the same waveform and channel realization but without additive noise. The receiver operator $\mathcal{R}_s\{\cdot\}$ applies timing synchronization, optional residual-Doppler compensation, and \ac{OFDM} demodulation, producing the grid $\mathbf Y^{(s)}=\mathcal{R}_s\{\mathbf y\}$. Using synchronization parameters inferred from $\mathbf y$, the matched noise-free effective target is
\begin{equation}
\mathbf{H}^{\mathrm{eff}}_s=
\mathcal{R}_s\{\mathbf{y}^{\mathrm{clean}}\}\oslash\mathbf{X},
\label{eq:effective_label}
\end{equation}
where $\mathbf X$ is the transmitted resource grid and $\oslash$ denotes element-wise division over its occupied \acp{RE}. The labels $\mathrm{clean}$ and $\mathrm{eff}$ denote the noise-free waveform and effective channel, respectively; $Y^{(s)}_{i,k}$ is entry $(i,k)$ of $\mathbf Y^{(s)}$. Unlike the canonical channel $\mathbf H^{\mathrm{can}}$ used as the target in~\cite{HELENA2025}, this effective target follows the receiver synchronization and compensation operations. Under residual \ac{ICI}, this effective channel also depends on the transmitted grid.

Let $\mathcal{P}$ denote the set of \ac{DM-RS} pilot positions. Because $X_{i,k}$ is known at these positions, the pilot-domain \ac{LS} estimate is $\hat H^{\mathrm{LS},(s)}_{i,k}=Y^{(s)}_{i,k}/X_{i,k}$ for $(i,k)\in\mathcal{P}$. A learned estimator receives this sparse grid, its \ac{LI}, or model-specific pilot/data-aided features, and implements
\begin{equation}
\hat{\mathbf H}^{\mathrm{eff}}_s=f_{\Theta_s}(\mathbf H_{\mathrm{in},s},\mathbf u_s),
\label{eq:dl_mapping_ntn}
\end{equation}
where $f_{\Theta_s}$ denotes the estimator with parameters $\Theta_s$ trained for condition $s$, $\mathbf H_{\mathrm{in},s}$ its channel input ($\mathrm{in}$), and $\mathbf u_s$ its auxiliary pilots, masks, or side information. The objective is to minimize effective-channel estimation error within the receiver's inference-latency budget.

\begin{figure}[t]
    \centering
    \begin{tikzpicture}[x=1cm,y=1cm,>=stealth,font=\scriptsize]
        \fill[green!48] (0,0) ellipse (2.65 and 0.43);
        \node[draw,circle,fill=yellow!75,inner sep=1.5pt] (ue) at (-1.55,0.05) {\acs{UE}};
        \node[draw,circle,fill=white,inner sep=1.5pt] (bc) at (0,0.05) {BC};
        \node[draw,rounded corners,fill=gray!12,inner sep=3pt] (sat) at (0.20,2.25) {\acs{LEO} satellite};
        \draw[->,thick,shorten >=9pt] (sat.south west) -- (ue.center) node[pos=0.54,left] {$f_{d,\mathrm{sat}}$};
        \draw[->,thick,shorten >=9pt] (sat.south) -- (bc.center) node[pos=0.52,right] {$f_{d,\mathrm{common}}$};
        \pgfmathanglebetweenpoints{\pgfpointanchor{ue}{center}}{\pgfpointanchor{sat}{south west}}
        \let\ueElevationAngle\pgfmathresult
        \pgfmathanglebetweenpoints{\pgfpointanchor{bc}{center}}{\pgfpointanchor{sat}{south}}
        \let\bcElevationAngle\pgfmathresult
        \begin{scope}[shift={(-1.55,0.05)}]
            \draw[thin] (0.28,0) -- (0.93,0);
            \draw[->] (0.50,0) arc[start angle=0,end angle=\ueElevationAngle,radius=0.50];
            \node at (0.78,0.25) {$\theta_{\mathrm{UE}}$};
        \end{scope}
        \begin{scope}[shift={(0,0.05)}]
            \draw[thin] (0.28,0) -- (0.93,0);
            \draw[->] (0.50,0) arc[start angle=0,end angle=\bcElevationAngle,radius=0.50];
            \node at (0.78,0.30) {$\theta_{\mathrm{BC}}$};
        \end{scope}
        \node at (0,-0.58) {Beam footprint};
    \end{tikzpicture}
    \caption{Geometry of beam-common Doppler pre-compensation.}
    \label{fig:ntn_system_model}
\end{figure}

The two receiver configurations share transmitter pre-compensation and joint \ac{PDSCH} timing--frequency alignment. \acs{NTN}-1 then estimates and compensates the residual Doppler $\hat f_{d,\mathrm{RX},1}$, whereas \acs{NTN}-2 sets $\hat f_{d,\mathrm{RX},2}=0$, leaving residual phase evolution in its input and effective target. Thus, \acs{NTN}-1 represents receiver-compensated operation, while \acs{NTN}-2 provides a controlled stress case without this correction. Matched channel and noise realizations isolate the effect of receiver Doppler processing.

%% file: sections/section3.tex
\section{Evaluation Methodology and Estimators}\label{sec:methodology}

All estimators output the full channel grid, with complex quantities represented by separate real and imaginary components. Learned models are trained independently for \acs{NTN}-1 and \acs{NTN}-2. To prevent data leakage, channel realizations are partitioned before generating their \ac{SNR} variants, ensuring that all noisy versions of a given realization belong to the same dataset split. Consequently, test samples are excluded from training, normalization, early stopping, and hyperparameter selection.

\subsection{Classical Estimators}\label{sec:classical_estimators}

The conventional references are \acf{LS} with \ac{2D} \acf{LI} (\acs{LS}+\acs{LI})~\cite{lee2008training}, a practical \emph{\ac{5G-NR} \ac{DM-RS} estimator}\cite{MathWorksNRChannelEstimate}, and three \ac{LMMSE} implementations~\cite{Edfors1998}. For either receiver condition, let $\mathbf h=\operatorname{vec}(\mathbf H_s^{\mathrm{eff}})$ be the vectorized effective target, $\mathbf h_p$ its pilot coefficients, and $\hat{\mathbf h}_p^{\mathrm{LS}}$ the corresponding noisy pilot observations. Here, $p$ denotes restriction to the pilot positions $\mathcal P$; the condition index $s$ is suppressed in this estimator for readability. The implemented affine \ac{LMMSE} estimator is
\begin{equation}
\hat{\mathbf h}=\overline{\mathbf h}+\mathbf R_{hp}
\left(\mathbf R_{pp}+(\hat\sigma^2+\lambda\alpha)\mathbf I\right)^{-1}
\left(\hat{\mathbf h}_p^{\mathrm{LS}}-\overline{\mathbf h}_p\right),
\label{eq:lmmse}
\end{equation}
where $N_p$ is the number of pilot coefficients. The overbars denote fitted means, $\mathbf R_{hp}$ is the centered sample cross-covariance between the full-grid target and its noise-free pilot coefficients, $\mathbf R_{pp}$ is the centered sample covariance of those pilot coefficients, and $\mathbf I$ is the $N_p\times N_p$ identity matrix. The fitting partition supplies the means and covariances and the \ac{SNR}-conditioned pilot-error variance $\hat\sigma^2$, measured between noisy pilot \ac{LS} observations and matched noise-free pilot labels. This fitted pilot-error variance is distinct from the additive-noise variance $\sigma^2$ in~\eqref{eq:received_ntn}. To improve numerical stability, we apply diagonal loading with $\lambda=10^{-4}$ and $\alpha=\max\{\mathrm{tr}(\mathbf R_{pp})/N_p, \varepsilon_{\mathrm{mach}}\}$.
Here, $\lambda$ controls the loading strength, while $\alpha$ scales it to the mean pilot-channel variance. The lower bound $\varepsilon_{\mathrm{mach}}$ is the single-precision machine epsilon, and $\mathrm{tr}(\cdot)$ denotes the matrix trace.

The \acs{NTN}-\ac{TDL} profiles A--D define different tap-delay and power structures, scaled by the separately varied delay spread~\cite{TR38811}. We evaluate three levels of statistical knowledge. Global \ac{LMMSE} fits one covariance across the training and validation profiles, combined with 11 \ac{SNR}-conditioned pilot-error variances. Profile-aware \ac{LMMSE} fits a covariance per profile, pooling its two delay spreads; the resulting 44 filters are selected using the true profile and nominal \ac{SNR}. Test-fitted profile-aware \ac{LMMSE} also uses test labels when fitting statistics, providing a non-deployable empirical linear reference. The practical \ac{NR} estimator uses \ac{CDM}-aware pilot processing with default adaptive averaging and full-grid interpolation~\cite{Li2000,Damjancevic2021}.

\subsection{DL Baselines}\label{sec:dl_baselines}

The learned baselines comprise \ac{SRCNN} and ChannelNet~\cite{soltani2019deep}, \ac{EDSR}~\cite{maruyama2021}, \ac{AttRNet}~\cite{Gao2025}, \ac{ProEsNet}~\cite{Zhang2024}, \ac{LSiDNN}-48~\cite{Sharma2024}, \ac{CEViT}~\cite{CEViT2023}, and \ac{MDELAN}-\ac{SISO}~\cite{MDELAN2026}. 

Compared with our terrestrial \acs{HELENA} study~\cite{HELENA2025}, we revise input processing, reconstruction, and selected components to follow the source architectures more closely. Table~\ref{tab:implementation_updates} summarizes the changes and inputs. ChannelNet retains its 20-layer \ac{DnCNN}, and \acs{EDSR} uses eight residual blocks; both use 32 feature filters instead of the original 64. ChannelNet retains the 64/32/1-filter \acs{SRCNN} front end.

\protect\acs{SRCNN} and ChannelNet receive interpolated \ac{LS} grids and process I/Q separately with shared weights; ChannelNet batches both components in one call. \Ac{EDSR} receives a half-resolution interpolated grid and reconstructs it through learned upsampling. In contrast, \ac{AttRNet}, \ac{ProEsNet}, and \ac{LSiDNN}-48 receive pilot-location \ac{LS} coefficients. \Ac{ProEsNet} reconstructs the grid progressively through three supervised stages and returns the final estimate during inference. \acs{CEViT} receives an interpolated \ac{LS} grid with nominal \ac{SNR}, mobility-related maximum Doppler, and configured delay spread. These configured scenario parameters are standardized using training-partition statistics. We use the CE-ViT architecture~\cite{CEViT2023} with fixed \ac{NR} pilots.

We adapt the dual-port uplink \ac{MDELAN} estimator \cite{MDELAN2026} to the common \ac{SISO} downlink grid using two blocks and the communication-informed preprocessing of MDX~\cite{MDX2025}. Interpolated \ac{PA-LS} and a practical noise estimate yield per-\ac{RE} \ac{SISO} \ac{MMSE} symbol estimates, from which \ac{DA-LS} features are constructed. The blocks process \ac{PA-LS}, \ac{DA-LS}, and positional inputs into the full channel estimate.

\subsection{HELENA}\label{sec:helena_models}

\begin{figure*}[t]
    \centering
    \includegraphics[width=\linewidth]{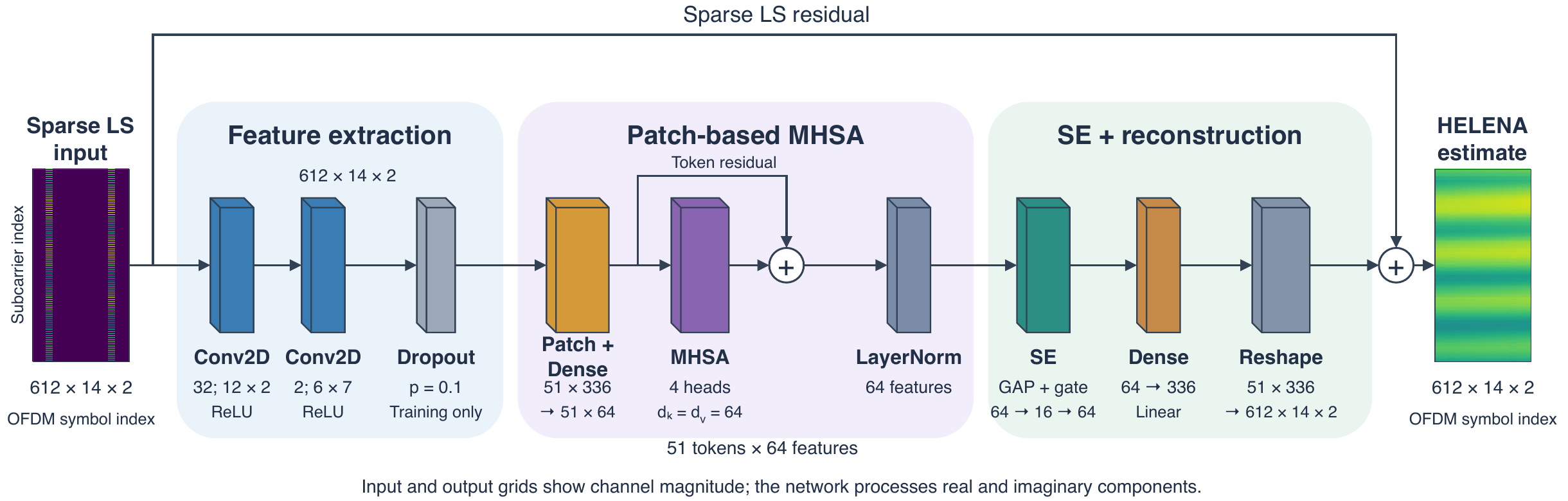}
    \caption{HELENA architecture with sparse input and predicted channel magnitude from one \acs{NTN}-1 test sample.}
    \label{fig:helena_architecture}
\end{figure*}

Fig.~\ref{fig:helena_architecture} summarizes \ac{HELENA}~\cite{HELENA2025}. Each Conv2D label gives the output-filter count followed by the kernel size in subcarriers $\times$ OFDM symbols: the two layers use 32 filters with $12\times2$ kernels and two filters with $6\times7$ kernels, respectively, both with \ac{ReLU}. The resulting features are partitioned into 51 non-overlapping frequency patches and projected into 64-dimensional tokens. \Ac{MHSA} models dependencies between these tokens using four heads, with per-head key and value dimensions $d_k=d_v=64$.

Following the attention residual addition and layer normalization, the \ac{SE} block uses \ac{GAP} across tokens. Two dense layers of widths 16 and 64 use \ac{ReLU} and sigmoid activations, respectively. The sigmoid outputs form a channel-wise gate that rescales each token's features. A linear head reconstructs the full grid, with a global residual connection adding the sparse \ac{LS} input to the learned correction. \acs{HELENA}-\acs{MHSA} removes the \acs{SE} block to isolate its contribution; both retain their terrestrial architectures.

In this work, we also introduce \acs{HELENA}-H4K32, a variant that reduces the per-head key and value dimensions from 64 to 32 while retaining four attention heads and the remainder of the architecture. With 83,138 parameters, compared with 116,290 for \acs{HELENA}, it enables us to evaluate the impact of reduced attention capacity on the accuracy--latency trade-off.

\begin{table*}[t]
\centering\scriptsize
\caption{Estimator inputs and implementation updates since the terrestrial \acs{HELENA} benchmark~\cite{HELENA2025}.}
\label{tab:implementation_updates}
\label{tab:model_summary}
\setlength{\tabcolsep}{2pt}
\renewcommand{\arraystretch}{1.12}
\begin{tabularx}{\textwidth}{@{}>{\raggedright\arraybackslash}p{1.4cm}*{11}{>{\raggedright\arraybackslash}X}@{}}
\toprule
 & \acs{SRCNN} & \shortstack{ChannelNet\\F32} & \shortstack{\acs{EDSR}\\B8/F32} & \acs{AttRNet} & \acs{ProEsNet} & \shortstack{\acs{LSiDNN}\\48} & \acs{CEViT} & \shortstack{\acs{MDELAN}\\\acs{SISO}} & \acs{HELENA} & \shortstack{\acs{HELENA}\\\acs{MHSA}} & \shortstack{\acs{HELENA}\\H4K32}\\
\midrule
Current input and auxiliary information & Interpo\-lated \acs{LS} & Interpo\-lated \acs{LS} & Half-resolution interpolated \acs{LS} & Pilot-location \acs{LS} & Pilot-location \acs{LS} & Pilot-location \acs{LS} & Interpo\-lated \acs{LS}; nominal \acs{SNR}, mobility Doppler, delay spread & \acs{PA-LS}, \acs{DA-LS}, position; practical noise estimate & Sparse \acs{LS} grid & Sparse \acs{LS} grid & Sparse \acs{LS} grid\\
\midrule
HELENA-TN implementation & Joint I/Q processing & Joint I/Q processing & Sixteen blocks; full-grid refinement & Full-grid input; two-channel attention bottleneck & Full-grid refinement; 64 feature channels & Flattened zero-filled full grid & Adapted patch projection; metadata tokens; two encoders & Not evaluated & Full-grid convolutions, \acs{MHSA}, \acs{SE}, and residual reconstruction & \acs{HELENA} without \acs{SE} & Not evaluated\\
\midrule
Updates in this paper & Separate I/Q processing with shared weights & Batch I/Q through shared \acs{SRCNN}--\acs{DnCNN}; retain depth and 32 \acs{DnCNN} filters & Eight blocks; residual scaling; $\times2$ pixel-shuffle upsampling; retain 32 filters & Pilot-grid input; 16-channel bottleneck; bilinear upsampling and $36\times7$ output convolution & Three-stage reconstruction; four feature channels; intermediate supervision & Pilot-only \acs{LS} vector; retain 48 hidden neurons & Explicit $10\times4$ patches; metadata feature columns; learned positional encoding; one encoder and inverse patch reconstruction & Add a two-block \acs{SISO} downlink estimator adapted from the dual-port uplink model; tune training hyperparameters & Un\-changed & Un\-changed & Reduce per-head key/value dimensions from 64 to 32; retain four heads\\
\bottomrule
\end{tabularx}
\end{table*}

%% file: sections/section5.tex
\section{NTN Dataset Creation}\label{sec:dataset}

The paired datasets implement the two receiver operators $\mathcal{R}_1$ and $\mathcal{R}_2$ defined in Section~\ref{sec:system}. They are generated at waveform level using MATLAB 5G Toolbox and Satellite Communications Toolbox, following the \ac{NR} \ac{NTN} \ac{PDSCH} chain\cite{MathWorksNTNPDSCH}. Table~\ref{tab:ntn_dataset} shows that propagation, numerology, pilots, and scenario sampling are common; receiver-side residual-Doppler compensation is the controlled difference.

\begin{table}[t]
\centering
\caption{Paired NTN dataset configuration.}
\label{tab:ntn_dataset}
\setlength{\tabcolsep}{3.5pt}
\renewcommand{\arraystretch}{1.04}
\resizebox{0.9\columnwidth}{!}{%
\begin{tabular}{lll}
\toprule
\textbf{Parameter} & \textbf{NTN-1} & \textbf{NTN-2} \\
\midrule
Carrier/grid & \multicolumn{2}{c}{2 GHz; 51 RB; 30 kHz SCS; normal CP} \\
Grid size & \multicolumn{2}{c}{$612\times14$ complex resource elements} \\
Link/antennas & \multicolumn{2}{c}{Downlink PDSCH; SISO} \\
Waveform symbols & \multicolumn{2}{c}{Full-grid QPSK} \\
DM-RS & \multicolumn{2}{c}{Type A; pos. 2; add. pos. 1; config. type 2} \\
Channel profiles & \multicolumn{2}{c}{NTN-TDL-A, B, C, and D} \\
Delay spread A/B & \multicolumn{2}{c}{30 or 100 ns} \\
Delay spread C/D & \multicolumn{2}{c}{5 or 30 ns} \\
Satellite altitude & \multicolumn{2}{c}{600 km} \\
UE elevation/speed & \multicolumn{2}{c}{10--89$^{\circ}$; 0--120 km/h} \\
Pre-RX residual Doppler & \multicolumn{2}{c}{150--900 Hz; beam offset up to 1.5$^{\circ}$} \\
SNR/examples & \multicolumn{2}{c}{0:2:20 dB; 11,264 per dataset} \\
PDSCH alignment & \multicolumn{2}{c}{Joint time--frequency} \\
TX pre-compensation & Enabled & Enabled \\
RX Doppler compensation & Independent time--frequency & Disabled \\
Primary target & Compensated & Residual-impaired \\
\bottomrule
\end{tabular}}
\end{table}

Each dataset contains 256 examples per \ac{SNR} and NTN-TDL profile, i.e., $4\times11\times256=11{,}264$ samples. Channel realizations are balanced across the A--D multipath families, their applicable delay spreads, and three \ac{UE}-speed, elevation, and residual-Doppler bins. Per profile, 179/38/39 realizations are assigned to training/validation/testing before generating their \ac{SNR} variants, yielding 7,876/1,672/1,716 samples. Identical waveform, channel, and noise seeds make the two receiver conditions sample-wise comparable. SNR denotes the nominal AWGN injection level after normalized waveform/channel scaling; the dataset is intended for estimator comparison rather than satellite link-budget evaluation.

Both conditions apply beam-common waveform pre-compensation to obtain $x_{\mathrm{pc}}(t)$ and estimate $\hat\tau_s$ through joint PDSCH frequency/timing alignment. NTN-1 estimates $\hat f_{d,\mathrm{RX},1}$ by combining cyclic-prefix fractional-offset and DM-RS integer-subcarrier estimates~\cite{vanDeBeek1997,Chen2025NTNDoppler}. NTN-2 retains the alignment and timing tracking but sets $\hat f_{d,\mathrm{RX},2}=0$, as in Section~II. All receiver-side timing and residual-Doppler estimates use noisy observations rather than true offsets. 

To avoid synchronization failures that would dominate the channel-estimation task, both conditions use the same offline quality filter. Candidate realizations whose correction-enabled reference retains an absolute residual frequency offset greater than half the subcarrier spacing are regenerated with new noise realizations while preserving sample pairing. Fewer than 0.3\% of samples required regeneration. Finally, the practical baseline uses \texttt{nrChannelEstimate} from \cite{MathWorksNRChannelEstimate}, configured with \ac{CDM} lengths, averaging, interpolation, and noise estimation.

Each MAT file stores the noisy grid, sparse/interpolated LS inputs, pilots and mask, practical DM-RS estimate, effective target, auxiliary canonical channel, split indices, SNR, scenario ID, and physical/receiver metadata. The evaluation assumes a known PDSCH/DM-RS configuration, isolating channel estimation from SSB detection, PBCH/control acquisition, and stateful initial synchronization.

%% file: sections/section6.tex
\section{Performance Evaluation}\label{results}

\subsection{Experimental Setup and Reproducibility}
The test partition contains 156 independent channel realizations (39 per profile), each evaluated at eleven \acp{SNR}, yielding 1,716 samples per receiver condition. Each learned model/condition uses one fixed-seed training run. Learned accuracy in Table~\ref{tab:accuracy_summary} uses RTX \ac{FP16}; classical estimators use native precision. \Ac{FLOPs} are whole-graph TensorFlow profiler counts at batch size one.

Learned models use Adam optimizer, batch size 64, at most 500 epochs, and early stopping with patience 50 and best-validation restoration. The initial learning rate is $10^{-3}$, reduced by 0.8 after 40 epochs without validation improvement, down to $10^{-5}$. The objective is \ac{MSE}; \ac{ProEsNet} sums three stage-wise losses. ChannelNet freezes the \ac{SRCNN} front end pretrained on the same condition. Training/validation tuning selects a learning rate of 0.03, no kernel $L_2$ regularization, and residual-scale initialization of $10^{-4}$ for \ac{MDELAN}-\ac{SISO}.

\begin{figure*}[t]
    \centering
    \includegraphics[width=0.95\linewidth]{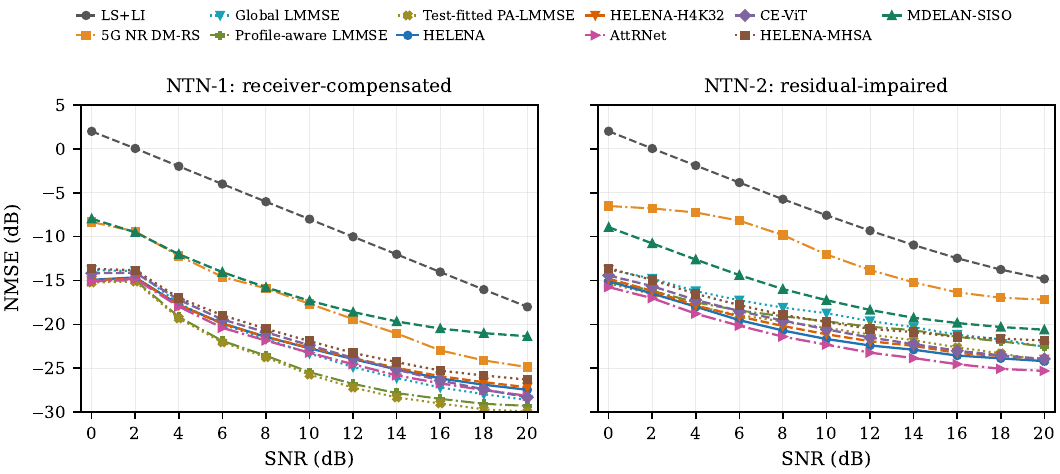}
    \caption{Test \acs{NMSE} versus \acs{SNR} for learned and classical estimators.}
    \label{fig:nmse_snr_ntn}
\end{figure*}

Keras models are exported to \ac{ONNX} and verified numerically before TensorRT compilation with a 4~GiB workspace and batch size one. The modeled downlink \ac{CE} is performed at the \ac{UE}; RTX PRO 4500 and 10~W Jetson Orin NX represent high-performance and embedded \ac{GPU} receiver platforms, respectively. Both use \ac{FP16}, with TensorRT~10.16 on RTX and 10.3 on Orin\footnote{\url{https://developer.nvidia.com/embedded/jetpack-sdk-62}}. RTX TensorRT \ac{FP16} inference preserves the TensorFlow \ac{FP32} reference accuracy: across eleven models and both
conditions, the absolute difference in SNR-averaged \ac{NMSE} is at most 0.0029~\acs{dB}. Datasets, models, and scripts will be released online \footnote{\url{https://github.com/miguelhdo/anna-helena-ntn}}.

Each model/platform uses 100 runs of 500 warm-up and 5,000 timed calls, with balanced randomized ordering and five-second idle guards. Percentiles pool the 500,000 calls; latency standard deviation is across the 100 run means. Prepared-input-to-channel-output latency includes host-to-device and device-to-host transfers, inference, and synchronization, excluding input construction and the remaining receiver chain. Energy is measured separately over ten sustained-workload runs, each lasting at least ten seconds and 5,000 calls after warm-up, with power sampled every 20~ms. Energy per sample is total workload energy divided by completed samples; its standard deviation is across the ten runs.

\input{tables/accuracy_summary}
\input{tables/accelerated_inference}

\begin{figure*}[t]
    \centering
    \includegraphics[width=0.9\linewidth]{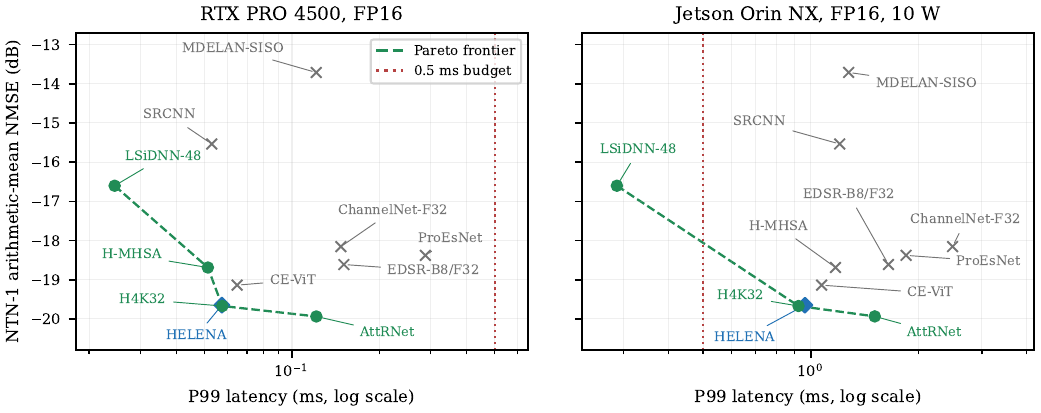}
    \caption{NTN-1 accuracy--latency trade-off. H4K32 and H-MHSA denote the HELENA variants.}
    \label{fig:pareto_ntn1}
\end{figure*}
\subsection{\acs{CE} Accuracy}

Fig.~\ref{fig:nmse_snr_ntn} shows that the \acs{NTN}-2 curves flatten earlier as \ac{AWGN} decreases and residual impairments become dominant. To summarize both overall accuracy and sensitivity to these impairments, Table~\ref{tab:accuracy_summary} reports the equally weighted arithmetic mean of the eleven per-\ac{SNR} linear \ac{NMSE} values over 0--20~\acs{dB}, converted to \acs{dB} after averaging. The paired differences, $\Delta_{\mathrm{all}}$ over 0--20~\acs{dB} and $\Delta_{\mathrm{H}}$ over 12--20~\acs{dB}, are \acs{NTN}-2 minus \acs{NTN}-1 using the same aggregation. Positive values indicate higher error with residual impairments; percentages compare the corresponding linear means.

Among the learned estimators, \ac{AttRNet} achieves the lowest observed mean \ac{NMSE}. \Ac{HELENA} remains within 0.29~\acs{dB} in \acs{NTN}-1 and 0.72~\acs{dB} in \acs{NTN}-2, while reducing error relative to \ac{CEViT} by 11.1/16.7\%. It also outperforms the adapted \ac{NTN}-specific \ac{MDELAN}-\ac{SISO}. The unchanged architecture therefore remains competitive with both terrestrial and \ac{NTN}-specific learned estimators. The deployment measurements in Section \ref{dep_per} will show that this performance is achieved with substantially lower latency and energy cost than \acs{AttRNet}.

The \ac{HELENA} variants help explain this accuracy. Relative to \acs{HELENA}-\ac{MHSA}, channel recalibration reduces error by 19.8/32.2\%, with the larger benefit under residual impairments. \acs{HELENA}-H4K32 nearly matches the original architecture in \acs{NTN}-1, within 0.023~\acs{dB}, but has 0.332~\acs{dB} higher error in \acs{NTN}-2. In the evaluated configuration, reduced attention capacity therefore preserves compensated-channel accuracy more closely than in the residual-impaired case.

The statistical baselines exhibit a condition-dependent trend. Under receiver compensation, \ac{HELENA} reduces error relative to global \ac{LMMSE} by 9.2\%, whereas profile-aware and test-fitted filtering achieve lower error using the additional information described in Section~\ref{sec:classical_estimators}. Under residual impairments, however, \ac{HELENA} outperforms all evaluated \ac{LMMSE} variants. The ranking therefore reverses under the evaluated synchronization residuals, highlighting \acs{HELENA}'s ability to learn channel reconstruction under residual impairments.

At high \ac{SNR}, residual synchronization impairments limit the gains obtained by reducing noise, increasing error for all learned estimators. This degradation is less apparent in the full-range average, which is dominated by larger low-\ac{SNR} errors. The high-\ac{SNR} penalties therefore complement the overall
ranking, showing that the best-performing learned estimators are less affected by residual impairments than the practical \ac{NR} estimator and all evaluated \ac{LMMSE} variants.

\subsection{Deployment Performance}\label{dep_per}

Table~\ref{tab:accelerated_inference} reports deployment measurements for \acs{NTN}-1, since both conditions share architectures and tensor dimensions. At 30~kHz \ac{SCS},
each slot lasts 0.5~ms. Following the three-\ac{TTI} receiver-processing allowance used in \ac{HELENA}~\cite{HELENA2025,Damjancevic2021}, we allocate one slot to \ac{CE}, adopting $T_{\max}=0.5$~ms as the inference budget.

On RTX, all evaluated learned estimators meet this \ac{P99} budget. \Ac{HELENA} reaches 0.0573~ms \ac{P99} and consumes 5.726~mJ per sample, 67.3\% less than \ac{AttRNet}, while retaining competitive accuracy. \Ac{CEViT} consumes slightly less energy than \ac{HELENA}, but has higher mean latency and estimation error.

On Orin, \ac{HELENA} reaches 0.9620~ms \ac{P99} and consumes 3.144~mJ per sample. Relative to \ac{AttRNet}, it achieves 1.97$\times$ the mean inference speed with
66.0\% less module energy, but exceeds the adopted budget. Only \ac{LSiDNN}-48 meets the target, at 0.2874~ms \ac{P99}, with 3.05~\acs{dB} higher mean \ac{NMSE}
than \ac{HELENA}. Energy comparisons apply within each platform because RTX and Orin measurements cover different power domains.

Simplifying \ac{HELENA} offers limited deployment gains: removing \ac{SE} modestly reduces Orin mean latency and energy at the expense of accuracy and tail latency, while
reducing attention capacity removes 28.5\% of parameters with little latency or energy benefit. Neither variant meets the embedded budget. Fig.~\ref{fig:pareto_ntn1} captures the resulting trade-off: \ac{HELENA} and both variants lie on the RTX frontier, whereas H4K32 joins \ac{LSiDNN}-48 and \ac{AttRNet} on the Orin frontier.

%% file: tables/accuracy_summary.tex
\begin{table}[t]
\centering\scriptsize
\caption{Arithmetic-mean NMSE and learned-model complexity.}
\label{tab:accuracy_summary}
\setlength{\tabcolsep}{2pt}
\renewcommand{\arraystretch}{1.14}
\begin{tabularx}{\linewidth}{@{}>{\raggedright\arraybackslash}Xrrrrrr@{}}
\toprule
Estimator & \shortstack{Params\\($\times10^3$)} & \shortstack{FLOPs\\($\times10^9$)} & \multicolumn{2}{c}{NMSE (dB)} & \multicolumn{2}{c}{Penalty (dB)}\\
\cmidrule(lr){4-5}\cmidrule(l){6-7}
 & & & NTN-1 & NTN-2 & $\Delta_{\mathrm{all}}$ & $\Delta_{\mathrm H}$\\
\midrule
\acs{LS}+\acs{LI} & --- & --- & \cellcolor{red!10}-4.106 & \cellcolor{red!10}-3.954 & +0.151 & +1.320\\
\acs{5G-NR} \acs{DM-RS} & --- & --- & \cellcolor{red!10}-14.072 & \cellcolor{red!10}-10.108 & +3.964 & +6.279\\
G-\acs{LMMSE} & --- & --- & \cellcolor{red!10}-19.230 & \cellcolor{red!10}-17.695 & +1.535 & +5.791\\
PA-\acs{LMMSE}$^*$ & --- & --- & \cellcolor{green!15}-20.581 & \cellcolor{red!10}-18.710 & +1.871 & +6.902\\
TF-PA-\acs{LMMSE}$^{*\dagger}$ & --- & --- & \cellcolor{green!15}-20.767 & \cellcolor{red!10}-19.329 & +1.438 & +6.238\\
\midrule
\acs{SRCNN} & 8.129 & 0.277 & \cellcolor{red!10}-15.537 & \cellcolor{red!10}-14.342 & +1.195 & +4.801\\
ChannelNet F32 & 177.506 & 6.012 & \cellcolor{red!10}-18.155 & \cellcolor{red!10}-16.698 & +1.457 & +6.151\\
\acs{EDSR} B8/F32 & 194.817 & 1.676 & \cellcolor{red!10}-18.612 & \cellcolor{red!10}-17.167 & +1.445 & +6.280\\
\acs{ProEsNet} & 4.012 & 0.010 & \cellcolor{red!10}-18.378 & \cellcolor{red!10}-17.950 & +0.428 & +2.303\\
\acs{AttRNet} & 103.186 & 0.345 & \cellcolor{green!15}\textbf{-19.937} & \cellcolor{green!15}\textbf{-20.377} & -0.440 & +2.028\\
\acs{LSiDNN}-48 & 878.880 & 0.002 & \cellcolor{red!10}-16.602 & \cellcolor{red!10}-14.644 & +1.958 & +4.456\\
\acs{CEViT} & 204.416 & 0.130 & \cellcolor{red!10}-19.138 & \cellcolor{red!10}-18.868 & +0.270 & +3.170\\
\acs{HELENA}-\acs{MHSA} & 114.162 & 0.084 & \cellcolor{red!10}-18.688 & \cellcolor{red!10}-17.970 & +0.718 & +3.616\\
\acs{HELENA} & 116.290 & 0.084 & \cellcolor{blue!10}-19.648 & \cellcolor{blue!10}-19.660 & -0.013 & +2.399\\
\acs{HELENA}-H4K32 & 83.138 & 0.081 & \cellcolor{green!15}-19.670 & \cellcolor{red!10}-19.329 & +0.341 & +2.555\\
\acs{MDELAN}-\acs{SISO} & 1.396 & 0.018 & \cellcolor{red!10}-13.714 & \cellcolor{red!10}-14.375 & -0.662 & +0.505\\
\bottomrule
\end{tabularx}
\parbox{\linewidth}{\scriptsize Green/red: lower/higher error than HELENA; blue: HELENA; bold: best learned estimator. Penalties are NTN-2 minus NTN-1 over 0--20 dB ($\Delta_{\mathrm{all}}$) and 12--20 dB ($\Delta_{\mathrm H}$). G/PA/TF: global/profile-aware/test-fitted LMMSE. $^*$Uses true profile and nominal SNR; $^\dagger$also fits statistics using test labels and is non-deployable.}
\end{table}

%% file: tables/accelerated_inference.tex
\begin{table*}[t]
\centering
\caption{NTN-1 FP16 latency and energy.}
\label{tab:accelerated_inference}
\renewcommand{\arraystretch}{1.15}
\setlength{\tabcolsep}{2.2pt}
\resizebox{0.9\linewidth}{!}{%
\begin{tabular}{@{}lrrrrrrrrrr@{}}
\toprule
\multirow{2}{*}{Estimator} & \multicolumn{5}{c}{RTX PRO 4500 Blackwell} & \multicolumn{5}{c}{Jetson Orin NX, 10 W}\\
\cmidrule(lr){2-6}\cmidrule(l){7-11}
 & \shortstack{Mean$\pm$SD\\(ms)} & \shortstack{P99\\(ms)} & \shortstack{Energy$\pm$SD\\(mJ/sample)} & \shortstack{$\Delta T_{\mathrm{HEL}}$\\(\%)} & \shortstack{$\Delta E_{\mathrm{HEL}}$\\(\%)} & \shortstack{Mean$\pm$SD\\(ms)} & \shortstack{P99\\(ms)} & \shortstack{Energy$\pm$SD\\(mJ/sample)} & \shortstack{$\Delta T_{\mathrm{HEL}}$\\(\%)} & \shortstack{$\Delta E_{\mathrm{HEL}}$\\(\%)}\\
\midrule
\acs{SRCNN} & $0.0480\pm0.0006$ & \cellcolor{green!15}0.0529 & $7.377\pm0.053$ & -8.4 & +28.8 & $0.7819\pm0.0174$ & \cellcolor{red!10}1.2027 & $5.499\pm0.023$ & +15.9 & +74.9\\
ChannelNet F32 & $0.1429\pm0.0004$ & \cellcolor{green!15}0.1474 & $28.220\pm0.094$ & +172.6 & +392.8 & $2.4120\pm0.0125$ & \cellcolor{red!10}2.4835 & $24.336\pm0.058$ & +257.4 & +674.0\\
\acs{EDSR} B8/F32 & $0.1436\pm0.0009$ & \cellcolor{green!15}0.1509 & $19.564\pm0.133$ & +174.0 & +241.7 & $1.5708\pm0.0048$ & \cellcolor{red!10}1.6453 & $11.798\pm0.041$ & +132.8 & +275.2\\
\acs{ProEsNet} & $0.2561\pm0.0036$ & \cellcolor{green!15}0.2879 & $21.446\pm0.138$ & +388.7 & +274.5 & $1.7467\pm0.0093$ & \cellcolor{red!10}1.8421 & $10.578\pm0.056$ & +158.8 & +236.4\\
\acs{AttRNet} & $0.1116\pm0.0012$ & \cellcolor{green!15}0.1213 & $17.492\pm0.086$ & +113.0 & +205.5 & $1.3318\pm0.0083$ & \cellcolor{red!10}1.5070 & $9.244\pm0.057$ & +97.3 & +194.0\\
\acs{LSiDNN}-48 & $0.0218\pm0.0003$ & \cellcolor{green!15}0.0245 & $2.390\pm0.028$ & -58.5 & -58.3 & $0.2578\pm0.0036$ & \cellcolor{green!15}0.2874 & $1.493\pm0.012$ & -61.8 & -52.5\\
\acs{CEViT} & $0.0578\pm0.0008$ & \cellcolor{green!15}0.0647 & $5.395\pm0.026$ & +10.2 & -5.8 & $0.7939\pm0.0674$ & \cellcolor{red!10}1.0711 & $3.419\pm0.091$ & +17.7 & +8.7\\
\acs{HELENA}-\acs{MHSA} & $0.0475\pm0.0004$ & \cellcolor{green!15}0.0513 & $5.298\pm0.022$ & -9.4 & -7.5 & $0.6517\pm0.0153$ & \cellcolor{red!10}1.1704 & $2.970\pm0.029$ & -3.4 & -5.5\\
\rowcolor{blue!10}
\acs{HELENA} & $0.0524\pm0.0005$ & \cellcolor{green!15}0.0573 & $5.726\pm0.039$ & +0.0 & +0.0 & $0.6748\pm0.0239$ & \cellcolor{red!10}0.9620 & $3.144\pm0.023$ & +0.0 & +0.0\\
\acs{HELENA}-H4K32 & $0.0516\pm0.0008$ & \cellcolor{green!15}0.0575 & $5.856\pm0.032$ & -1.5 & +2.3 & $0.6726\pm0.0198$ & \cellcolor{red!10}0.9233 & $3.120\pm0.020$ & -0.3 & -0.8\\
\acs{MDELAN}-\acs{SISO} & $0.1165\pm0.0003$ & \cellcolor{green!15}0.1212 & $10.066\pm0.036$ & +122.2 & +75.8 & $0.9343\pm0.0092$ & \cellcolor{red!10}1.2739 & $6.160\pm0.013$ & +38.5 & +95.9\\
\bottomrule
\end{tabular}}
\parbox{\linewidth}{\scriptsize SD is across 100 latency-run means or ten energy runs. Green/red P99 cells meet/exceed 0.5 ms. Energy includes idle consumption and covers the RTX GPU board or Orin module input; compare within each platform. Positive $\Delta T_{\mathrm{HEL}}$/$\Delta E_{\mathrm{HEL}}$ indicates higher mean latency/energy than HELENA.}
\end{table*}

%% file: sections/section7.tex
\section{Discussion and Practical Implications}\label{sec:discussion}
The results support adapting terrestrial \ac{CE} architectures to \acp{NTN} through retraining without an \ac{NTN}-specific redesign. The preferred estimator depends on receiver constraints: \ac{AttRNet} achieves the lowest observed estimation error, whereas \ac{HELENA} offers a more favorable accuracy--latency--energy trade-off without requiring profile labels or channel statistics during inference. Under receiver
compensation, profile-aware \ac{LMMSE} achieves lower error using the true channel profile and nominal \ac{SNR}, information that may be unavailable in deployment. Under residual impairments, however, \ac{HELENA} outperforms
all evaluated \ac{LMMSE} variants. These findings support supervised adaptation of terrestrial architectures to \ac{NTN} conditions.

The results also show that receiver-side Doppler compensation remains beneficial even when learning-based estimation is employed. On the latency part, embedded deployment is considerably more restrictive than inference on a high-performance \acs{GPU}: for the evaluated Orin platform, only \ac{LSiDNN}-48 satisfies the adopted 0.5~ms \ac{P99} target. Reducing attention capacity preserves most of \ac{HELENA}'s accuracy but provides only modest latency and energy benefits. Consistent with~\cite{HELENA2025}, analytical complexity alone is therefore a poor predictor of deployed performance, making target-platform measurements essential for model selection. 

The present analysis is limited to the evaluated \ac{SISO}, 600-km \ac{LEO}, 2-GHz carrier, 30-kHz \ac{SCS}, and \ac{DM-RS} configuration. Link-level evaluation is still required to determine how the observed \ac{NMSE} gains translate into \ac{BLER} and throughput performance under practical operating conditions.

%% file: sections/section8.tex
\section{Conclusion and Future Work}\label{sec:conclusion}
This work evaluated whether terrestrial \ac{DL}-based channel estimators remain effective in \acp{NTN} after retraining. Using paired receiver-compensated and residual-impaired datasets, we compared eleven learned configurations and classical estimators under common \ac{5G-NR} \ac{NTN} conditions. The results show that terrestrial architectures can transfer successfully to \acp{NTN} without architectural modification, supporting retraining as a viable adaptation strategy. Among the evaluated methods, \ac{HELENA} achieved a favorable balance between channel-estimation accuracy, latency, and energy consumption. While more accurate learned estimators exist, their additional complexity resulted in less attractive deployment characteristics. The results further show that receiver-side Doppler compensation remains important and that achieving accurate low-latency embedded estimation is a more difficult challenge than achieving high estimation accuracy alone. 

Future work will evaluate link-level \ac{BER}, \ac{BLER}, throughput, and \ac{MIMO} performance, investigate terrestrial-to-\ac{NTN} zero-shot transfer and/or transfer learning and low-precision deployment techniques such as \ac{INT8} quantization, and explore both novel architectures and platform-specific optimizations for low-latency embedded \ac{NTN} receivers.